\documentclass{PoS}
\usepackage{amssymb,amsmath,multirow,epsfig,graphicx,color}

\usepackage{epsfig}
\usepackage{graphicx,amsmath}
\usepackage{color}
\usepackage{booktabs}
\usepackage{axodraw2}
\newcommand\ba{\begin{eqnarray}}
\newcommand\ea{\end{eqnarray}}

\newcommand{\be}{\begin{equation}}
\newcommand{\ee}{\end{equation}}
\newcommand{\bas}{\begin{eqnarray*}}
	\newcommand{\eas}{\end{eqnarray*}}

\newcommand{\pslash}{\slash \hspace{-0.26cm} P}
\newcommand{\kslash}{\slash \hspace{-0.22cm} k}

\def\psla{ \rlap \slash \! }  
\def\auj{\number\day\space\ifcase\month\or
	janvier\or f\' evrier\or mars\or avril
	\or mai\or juin\orjuillet\or ao\^ut
	\or septembre\or octobre\or novembre
	\or d\' ecembre\fi\space\number\year}  
\def\hoje{\number\day\space de \ifcase\month\or
	Janeiro,\or Fevereiro,\or Mar\cc o,\or 
	Abril,\or Maio,\or Junho,\or Julho,
	\or Agosto,\or Setembro,\or Outubro,\or Novembro,
	\or Dezembro,\fi\space\number\year}
\def\data{\number\day\space  \ifcase\month\or
	January,\or February,\or March,\or 
	April,\or May,\or June,\or July,
	\or Agust,\or septembre,\or Octubre,\or November,
	\or December,\fi\space\number\year}

\title{Pion observables with the Minkowski Space Pion Model  
\thanks{
	This work was supported in part by CAPES, and 
	Conselho Nacional de Desenvolvimento Cient\'ifico e Tecnol\'ogico (CNPq) 
	under grants 308025/2015-6 (JPBCM), 308486/2015-3 (TF).
	Funda\c{c}\~ao de Amparo \`a Pesquisa do Estado de S\~ao Paulo
	(FAPESP) under the thematic projects 2013/26258-4 and 2017/05660-0, 
	and by regular project 2019/02923-5 (JPBCM). Project INCT-FNA Proc. No. 464898/2014-5.
   }}
	
  \ShortTitle{Minkowski Space Pion Model}

\author{\speaker{J. P. B. C. de Melo}\\  \thanks{ LFTC-19-14/52 } 
    Laborat\'orio de F\'\i sica Te\'orica e 
     	Computacional - LFTC \\
     	Universidade Cruzeiro do Sul~/~Universidade Cidade de S\~ao Paulo, 
     	01506-000 S\~ao Paulo, Brazil  \\
        E-mail: \email{joao.mello@cruzeirodosul.edu.br}}

\author{R\^omulo M. Moita  \\
	Instituto Tecnol\'ogico de
		Aeron\'autica, DCTA \\ 12.228-900 S\~ao Jos\'e dos Campos, SP,
		Brazil.
	  \\
	E-mail: \email{rdmoyses@hotmail.com}}

\author{T. Frederico \\
Instituto Tecnol\'ogico de
	Aeron\'autica, DCTA \\ 12.228-900 S\~ao Jos\'e dos Campos, SP,
	Brazil. \\ 
	E-mail: \email{tobias@ita.br}
}

\abstract{
	The pion structure in Minkowski space is  described in 
	terms of an analytic model of the Bethe~-~Salpeter 	amplitude combined with Euclidean Lattice QCD results
	for the running quark mass. 
	In the present work, a pion model previously proposed, which
	allows for a Nakanishi integral representation, is studied in order to verify the sensitivity of the pion electromagnetic form factor
	to small variations of the quark self-energy. 
	In addition,  we 
	extend the previous work, providing the Nakanishi integral representation for the invariants associated with a
	 decomposition of the pion Bethe-Salpeter amplitude.	}

\FullConference{Light Cone 2019 - QCD on the light cone: from hadrons to heavy ions - LC2019\\
		16-20 September 2019\\
		Ecole Polytechnique, Palaiseau, France}

\begin{document}

\section{Introduction} 
In the present work, we extend the previous study performed  in~\cite{Clayton}
 to test the sensitivity of the pion observables to  the model parameters, and, 
check the limitations of the model presented with the original parameters.
The model is built  to fit the quark propagator in the space-like region obtained by
 Lattice QCD calculations in the 
Landau gauge~(see the reference~\cite{Clayton} for details), also  the  analytical model 
preserves the Lorentz invariance. The results from Lattice calculations used here, 
have two degenerate light quarks, u and d, and, the heavy quark s~\cite{Rojas2013,Parappilly2005}.  

The  quark model propagator is given by 
$S_F(k)=\imath\,Z(k^2)\left[\kslash-M(k^2)+\imath\epsilon\right]^{-1} $. Using that
the pion is very close to the chiral 
limit, as a simplification, we not considere the momentum 
dependence of the quark wave function renormalization factor, 
i.e, $Z(k^2)=1$.  Then, the model for the dressed quark propagator is written as $
S_F(k)=\imath 
(\kslash + M(k^2))\left( k^2-M^2(k^2)+\imath\epsilon \right)^{-1}.
$

The running quark mass model in the space-like region fits   
lattice calculations~\cite{Clayton,Rojas2013,Parappilly2005}, and
it is 
parametrized by the expression,
\begin{equation}
\label{runningmass}
M(k^2)=m_0-m^3\left[k^2- \lambda^2 +i \epsilon \right]^{-1}\, ,
\end{equation}
where 
$
m_0\,=\,0.014 \,\text{GeV},\,\, m\,=\,0.574\,\,\text{GeV}\,\, \text{and}
\,\,\lambda\,=\,0.846\,\text{GeV}$, which we name  initial 
parameter set~(IP)~\cite{Clayton}.  In the chiral 
limit, where the current quark mass vanishes, the scalar part of the 
self-energy  gives the invariant associated with the pseudoscalar component of the 
pion-quark-antiquark vertex. In this way, the present model 
incorporates effects from quark dressing and dynamical chiral symmetry 
breaking.

The quark propagator  can be written  in a factorized form, after solving $m^2_i=M^2(m^2_i)$:
\begin{equation}
S_F(k)=\imath\,\,\frac{\left(k^2- \lambda^2\right)^2\, 
	(\kslash +m_0)- \left(k^2- \lambda^2\right)\,
	m^3}
{{\prod_{i=1,3}}(k^2-m^2_i+\imath \epsilon)} \, ,
\label{sf1}
\end{equation}
where for the parameter set given above, only real poles at the positions,  
$$
m_1=0.371~GeV, \,\,\,\,\,\,\,\ m_2=0.644~GeV, \,\,\,\ \text{and}\,\,\,\,\ m_3= 0.954~GeV ,
$$
are found.
\begin{figure}[htb]
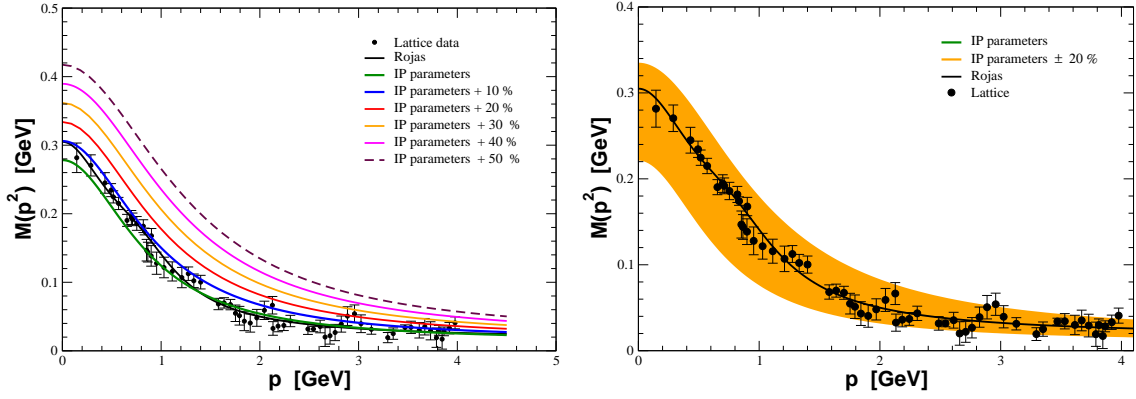

	\begin{center}
		\hspace*{-1.0024cm}
		\mbox{
			\epsfig{figure=massav2.eps ,width=7.30040cm ,angle=0}
			\hspace{0.0801003cm}
			\epsfig{figure=massav6.eps,width=7.30040cm,angle=0}
		} \par
		\caption{The running quark mass as a function of the momentum $p$, with 
			the parameters from the previous work~\cite{Clayton} compared to the results from
			the  parameters variations.  Also, in the figure are shown the LQCD  results~\cite{Parappilly2005} and
			and the parametrization given in Ref.\cite{Rojas2013} .
		}
		\label{fig1}
	\end{center} 
\end{figure}
The propagator in the form
\begin{equation}
\label{sfaltern}
S_F(k)=\imath\,\left[A(k^2)\,\kslash+B(k^2)\right] \ . 
\end{equation}
has for the self-energies:
\begin{flalign}
A(k^2)= \dfrac{\left(k^2- \lambda^2\right)^2}{{\prod_{i=1,3}}(k^2-m^2_i+\imath \epsilon)} 
, \ \ B(k^2)=\dfrac{(\lambda^2-k^2)m^3}{{\prod_{i=1,3}}(k^2-m^2_i+\imath \epsilon)}\,+\,A(k^2)m_0,
\label{defAB}
\end{flalign}
We can make for $A(k^2)$, the decomposition below, 
\begin{flalign} 
\dfrac{\left(k^2- \lambda^2\right)^2}{{\prod_{i=1,3}}(k^2-m^2_i)} 
&=\sum_{i=1}^{3} \frac{D_i}{\left(k^2-m_i^2\right)}\, ,
\end{flalign}
and  solving 
for $D's$ with the IP set, one gets: 
\begin{equation}
\label{valorD}
D_1=1.4992  ,\,\,\,\,\,\,\,\  D_2=-0.594098 \,\,\,\,\,\, 
\text{and}\,\,\,\,\,\,\,\ D_3=-0.0949811.
\end{equation}
Now, we must decompose $B(k^2)$ from Eq.(\ref{defAB}) in the same way as above,
\begin{flalign}
\frac{k^2m^3- \lambda^2m^3}{{\prod_{i=1,3}}(k^2-m^2_i)} + 
A(k^2) m_0 
&=\sum_{i=1}^{3} \frac{D_i m_0-E_i}{\left(k^2-m_i^2\right)}\, ,
\end{flalign}
and  we find the following solution for the $E's$: 
$$ 
\label{valorE}
E_1=0.42401285 ,\,\,\,\,\, E_2=-0.331377 \ \ \ \  \text{and}\,\,\,\,\,\,\,\ 
E_3=-0.07863548 \ .
$$ 
The spectral decomposition for $A(k^2)$ and $B(k^2)$ reads, 
\begin{eqnarray} 
\label{specrep}
A(k^2) =  \int_0^\infty d\mu^2 \frac{\rho_A(\mu^2)}{k^2-\mu^2+\imath\varepsilon} 
, \ \ B(k^2)& = & \int_0^\infty d\mu^2 \frac{\rho_B(\mu^2)}{k^2-\mu^2+\imath\varepsilon} \ , 
\end{eqnarray}
with the  spectral densities given by, 
$$
\rho_A(\mu^2)=-\frac1\pi \, \text{Im\,}[A(\mu^2)]\,\, \,\,\text{and}\,\,\,\,\rho_B(\mu^2)=-\frac1\pi \, \text{Im\,}[B(\mu^2)]
$$ 
which,  in principle, for a non-confining theory should  obey the positivity constraints for the  K\"allen-Lehman (KL) representation, 
$ 
\label{posab}
\mathcal{P}_a=\rho_A(\mu^2)\geq 0 \,\, \,\,\text{and}
\,\,\,\,\mathcal{P}_b=\mu\,\rho_A(\mu^2)-\rho_B(\mu^2)\geq 0 \ . $ 

We can write the function~$A(k^2)$,~as, 
\begin{flalign}
\int_0^\infty d\mu^2 \frac{\rho_A(\mu^2)}{k^2-\mu^2} &= \sum_{i=1}^3\dfrac{D_i}{k^2-m_i^2}=
\sum_{i=1}^3~ \int_0^\infty d\mu^2~\frac{D_i~\delta(\mu^2-m_i^2)}{k^2-\mu^2}
\end{flalign}
which leads to:
\begin{equation}\label{rhoA}
\rho_A(\mu^2)~=~D_1\delta(\mu^2-m_1^2)+D_2\delta(\mu^2-m_2^2)+D_3\delta(\mu^2-m_3^2).
\end{equation}
and also
\begin{flalign}\label{rhoB0}
\int_0^\infty d\mu^2 \frac{\rho_B(\mu^2)}{k^2-\mu^2} &= \sum_{i=1}^3\dfrac{E_i}{k^2-m_i^2}=
\sum_{i=1}^3~ \int_0^\infty d\mu^2~\frac{E_i~\delta(\mu^2-m_i^2)}{k^2-\mu^2} \ ,
\end{flalign}
which has a spectral density given by:
\begin{equation}\label{rhoB}
\rho_B(\mu^2)~=~E_1\delta(\mu^2-m_1^2)+E_2\delta(\mu^2-m_2^2)+E_3\delta(\mu^2-m_3^2)~+~m_0\rho_A(\mu^2) 
\ .
\end{equation}
We observe that the parametrization of the quark self energy leads to
a violation of the positivity constraints. 

The pion Bethe-Salpeter amplitude~(BSA) of this model can be written  in terms of the  Nakanishi integral 
representation~(NIR)~\cite{Nakanishi1963,Carbonell2010}. 
The pion-quark-antiquark vertex  denoted by~$\Gamma_\pi(k,P)$ 
has the general form below, 
\begin{equation}
\label{vertex}
\Gamma_\pi (k;P) = \gamma_5 [\imath E_\pi (k;P)+ \pslash  F_\pi (k;P)  
+ k^\mu P_\mu \ \kslash G_\pi (k;P) + \sigma_{\mu\nu} k^\mu P^\nu H_\pi (k;P)] \ , 
\end{equation}
Considering the chiral limit, we have for BSA model the structure below: 
\begin{equation}
\label{BSA1}
\Psi_\pi (k;P)=-\left[A(k_q^2)\,\psla{k}_q+B(k_q^2)\right]
\, \frac{\mathcal{N}\,\gamma_5 m^3 }{k^2-\lambda^2+\imath\epsilon} \,
\left[A(k_{\overline q}^2)\,\psla{k}_{\overline q}+B(k_{\overline q}^2)\right] \, 
\end{equation}
where $k_q=(k+P/2)$,  $k_{\overline q}=(k-P/2)$ and  Eq.(\ref{sfaltern}) for the quark propagator.
In order to obtain the integral representation of the BSA  model, we use Feynman's parameterization, elaborated in
 the identity below:
\begin{flalign}
	\label{ident}
\frac{1}{[(k+\frac{p}{2})^2-\mu^{\prime 2}+\imath\epsilon][k^2-\lambda^2+\imath\epsilon]
[(k-\frac{p}{2})^2-\mu^2+\imath\epsilon]}= \nonumber \\
=\int_{0}^{\infty} d\gamma\int_{-1}^1dz\frac{\mathcal{F}(\gamma,z~;\mu^{\prime },\mu,M)}
{\left[k^2+z\, k \cdot P+\gamma+\imath\epsilon\right]^3},
\end{flalign}
where 
\begin{equation}\label{functiong}
 \mathcal{F}(\gamma,z~;\mu^{\prime },\mu)~\equiv~
\frac{2~\theta(1+z-2\alpha)~\theta(\alpha-z)~\theta(1-\alpha)~ \theta(\alpha)}
{|2\lambda^2+M^2/4-\mu^{\prime^2}-\mu^{2}|},
\end{equation} 
and
\begin{equation}
\alpha(\gamma,z~;\mu^{\prime },\mu)=
\frac{\gamma-z(\mu^{2}-\lambda^2-M^2/4)+\lambda^2}{2\lambda^2+M^2/4-\mu^{2}-\mu^{\prime^2}}.
\end{equation}
The BSA from Eq.(\ref{BSA1}) can be decomposed in terms of the Dirac operators, 
\begin{flalign}
&\Psi_\pi (k;P)=\gamma_5\,\chi_1(k , P)+\psla{k}_q\,\gamma_5\,\chi_2(k,P)+
\gamma_5\,\psla{k}_{\overline q}\,\chi_3(k,P)+
 \psla{k}_q\,\gamma_5\,\psla{k}_{\overline q}\,\chi_4(k,P)~=~ 
\nonumber\\
&=-A(k_q^2)\,\psla{k}_q~ \frac{m^3\mathcal{N}\,\gamma_5}{k^2-\lambda^2+\imath\epsilon}~
A(k_{\overline q}^2)\,\psla{k}_{\overline q}-
                A(k_q^2)\,\psla{k}_q~ \frac{m^3\mathcal{N}\,\gamma_5}{k^2-\lambda^2+\imath\epsilon}~
                B(k_{\overline q}^2)\nonumber\\
                &~~~~-B(k_q^2)~ \frac{m^3\mathcal{N}\,\gamma_5}{k^2-\lambda^2+\imath\epsilon}~
                A(k_{\overline q}^2)\,\psla{k}_{\overline q}-
                B(k_q^2)~ \frac{m^3\mathcal{N}\,\gamma_5}{k^2-\lambda^2+\imath\epsilon}~
                B(k_{\overline q}^2),
                \label{sistema0}
\end{flalign}
and the invariants $\chi_i(k, P)$ can be written with the  Nakanishi integral representation (NIR), with weight functions
determined analytically as shown in the following.

In order to obtain the  invariants, $\chi_i(k,P)$, we introduce $A(k_q^2)$, $B(k_q^2)$,
$A(k_{\overline q}^2)$ and $B(k_{\overline q}^2)$ in Eq.(\ref{sistema0}), that leads to:
\begin{equation}
\chi_i(k,P)=
\int_0^\infty d\mu^{\prime 2} 
\frac{\rho_{(A,B)}(\mu^{\prime 2})}{[(k+\frac{P}{2})^2-\mu^{\prime 2}+\imath\varepsilon]}
\frac{m^3(\mathcal{-N})}{[k^2-\lambda^2+\imath\epsilon]}
\int_0^\infty d\mu^2
\frac{\rho_{(A,B)}(\mu^2)}
{[(k-\frac{P}{2})^2-\mu^2+\imath\varepsilon]}.
\end{equation}
One can write  the  scalar functions,~$\chi_i(k,p)$, with NIR as:
\begin{equation}
\chi_i(k,P)=\int_{-\infty}^{+\infty} d\gamma\int_{-1}^1dz \frac{{G}_i(\gamma,z;M)}
{\left[k^2+z\, k\cdot P+\gamma+\imath\epsilon\right]^3},
\label{chi1}
\end{equation}
and after replacing the spectral densities,~$\rho_{A}$~and~$\rho_{B} $, 
 and integrating over the Dirac delta's in the~$\chi_{i}$'s, 
we obtain  the weight functions:
\begin{small}
\begin{eqnarray}
{G}_1(\gamma,z;P)&=&-m^3
\mathcal{N}\sum\limits_{i,j}
(E_i+m_0D_i)(E_j+m_0D_j)F_{ij},\,\,\,
{G}_2(\gamma,z;M)=-m^3\mathcal{N}\sum\limits_{i,j}D_i~(E_j+m_0D_j)~F_{ij}, \nonumber
\\ 
{G}_3(\gamma,z;M)&=&-m^3\mathcal{N}\sum\limits_{i=1}^3\sum\limits_{j=1}^3~(E_i+m_0D_i)~D_j~F_{ij},
\,\,\,
{G}_4(\gamma,z;M)=-m^3\mathcal{N}\sum\limits_{i=1}^3\sum\limits_{j=1}^3~D_i~D_j~F_{ij}\, ,
\label{Gs}
\end{eqnarray}
\end{small}
where $1\leq i,\leq3j$ and  $F_{ij}$ are lengthy functions  computed with the help of  Eq. (\ref{functiong}), which will be presented elsewhere.

The light-front projection of  Eq. (\ref {chi1}) is the basic ingredient to obtain the valence wave function:
\begin{equation}
 \Psi_i(z,\vec{k}_{\perp })=
 \frac{P^+}{\sqrt2}~z(1-z)~\int_{-\infty}^{+\infty}\frac{dk^-}{2\pi}~\chi_i(k,P)=\frac{i}{8\sqrt{2}}(1-{z}^2)\int_{-\infty}^{\infty} d\gamma~
\frac{G_i(\gamma,z,P)}{[-{z}^2\frac{M^2}{4}-|\vec{k}_{\perp}|^2-\gamma]^2},,
\label{wfdef}
\end{equation}
where $z=-\frac{2k^+}{P^+}$ and we have choosen~$\vec{P}_{\perp}~=~0$.

In relation to the the previous work~\cite{Clayton}, 
we allowed a variation of the model parameters, to check  the robustness of the predictions for the
the pion electromagnetic form factor allowing some change in the quark mass function as the basic input from LQCD
calculations. Such variations gives also an idea of what to expect in terms of theory uncertainties when comparing with the forthcoming data from 
the TJLAB laboratory
(12~GeV upgrade TJLAB) for energies above  their first
results~\cite{Dudek2012}.
\begin{figure}[htb]
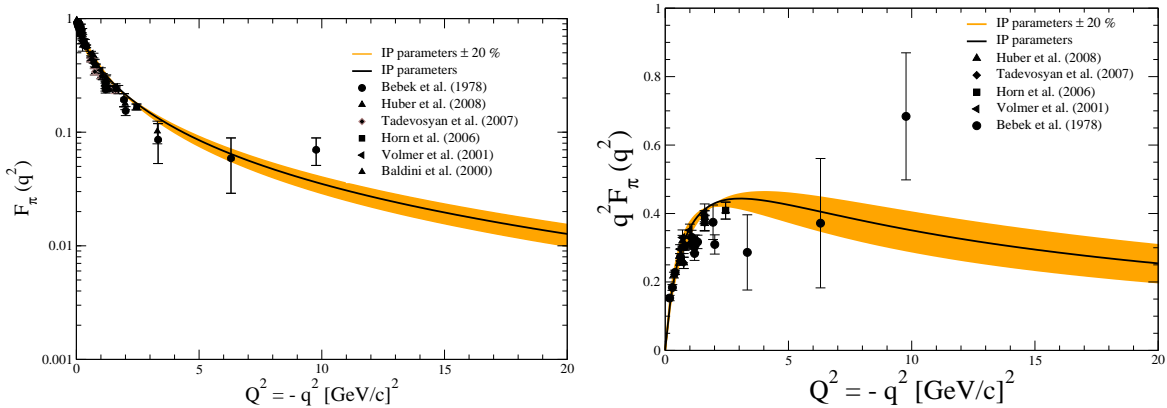

	\begin{center}
		\hspace*{-1.0024cm}
		\mbox{
			\epsfig{figure=ffpionv6.eps ,width=7.50040cm ,angle=0}
			\hspace{0.0801003cm}
			\epsfig{figure=q2ffpionv6.eps,width=7.50040cm,angle=0}
		} \par
		\caption{
			The pion electromagnetic form factor calculated with the model from~\cite{Clayton}, compared 
			with the experimental data. The band represents results obtained from a $\pm20 ~\%$ variation of the  parameters.  
		}
		\label{fig3}
	\end{center} 
\end{figure}

In the present work,  the pion space-like electromagnetic form factor is calculated
with a quark electromagnetic current operator that satisfies 
the Ward-Takahashi identity to ensure current conservation  \cite{Clayton}. For the original set of parameters 
we have a good agreement with the experimental electromagnetic radius for the pion,
$r^{Exp.}_{\pi}= 0.659\pm 0.004~[fm]$, and also, 
for  the weak pion decay constant, 
$f^{Exp.}_\pi = 90.276\pm0.0707 [MeV]$ (PDG \cite{PDG2014}).  
The new results for the pion electromagnetic form factor shown in Fig.(\ref{fig3}), electromagnetic radius and the weak decay constant, 
are found to be consistent with the experimental 
data \cite{Volmer2000,Horn2006,Tadevosyan2007,Huber2008,PDG2014}, taking into account
a  20 $\%$ variation for the model parameters. The resulting band encodes the present experimental data and provides an estimation of the  expected
error in the prediction of the form factor for large momentum transfers.
Also, in the present work, we sketch the derivation of  the Nakanishi weight functions of the model, which will allow to investigate the effect of the quark 
self-energy in these functions preparing the basis for more refined approaches.


\end{document}